\documentclass[aps,prl,twocolumn,amsmath,amssymb,superscriptaddress]{revtex4-1}
\usepackage{dcolumn}
\usepackage{bm}

\usepackage{color} 
\usepackage{ulem}

\usepackage{amsmath}
\usepackage{graphicx}
\usepackage{float}
\usepackage{subfig}
\usepackage{tikz}
\usepackage{color}
\usepackage[colorlinks,bookmarks=false,citecolor=blue,linkcolor=red,urlcolor=blue]{hyperref}

\definecolor{darkred}{rgb}{0.7,0.0,0.0}

\definecolor{darkblue}{rgb}{0,0.02,0.45}

\definecolor{darkgreen}{rgb}{0.02,0.45,0.0}

\definecolor{violet}{rgb}{0.8,0.2,0.6}

\providecommand{\U}[1]{\protect\rule{.1in}{.1in}}

\begin{document}

\title{Multipole phases in a type of spin ladders with local conserved quantities and generalizations}

\author{Jianlong Fu}
\affiliation{Department of Physics, Hong Kong University of Science and Technology, Clear Water Bay, Hong Kong, China}
\affiliation{Center for Theoretical Condensed Matter Physics, Hong Kong University of Science and Technology, Clear Water Bay, Hong Kong, China}

\begin{abstract}
We study spin ladder models with exact multipole phases, which are traditional spin phases formed by multipole moments. These phases feature non-trivial order with zero magnetization. The multipole models have dimer local conserved quantities that are Ising terms of spin. The Hilbert spaces are locally fragmented into independent sectors described effectively by ``higher-order spin". For dipole models, we consider two ladder geometries with quadratic spin couplings and work out the phase diagrams. Higher-order multipole models are obtained by introducing more dimer conserved quantities. The phases are characterized by the values of the local conserved quantities and the traditional spin phases of the higher-order spin. The dipole models can in principle be realized in experiments, we propose such realization by electric field controlled Rydberg atom arrays.
\end{abstract}

\maketitle

\section{Introduction}

The search for exotic phases with unconventional excitations has long been the focus of condensed matter physics. Besides anyons with nontrivial statistics \cite{Wilczek82,Kitaev06,Nayak08}, recently there has been growing interest in {\it multipoles}, which are nonlocal composite objects formed by opposite charges or spins aligned in specific order. In particular, studies on dimer Mott insulators $\kappa-(\text{ET})_{2}\text{Cu}_{2}(\text{CN})_{3}$ \cite{Hotta10,Naka10,Naka2016,Naka16,Yamada20} show that elementary excitations in dimerized systems can sometimes be described approximately by dipoles. Altermagnetism \cite{libor221,libor222}, a recently discovered magnetic phase, can be described by multipole order parameters \cite{Bhowal24,McClarty24}. To further understand how multipole excitations can emerge, it is interesting to look for phases of matter whose elementary degree of freedom (DOF) is multipoles. Such phases can possess non-trivial order formed by the multipole moments while the net magnetization remains zero; low-energy excitations are formed and described by the flipping of multipole moments. In this work, we demonstrate the existence of such {\it exact} multipole phases by constructing solvable parent Hamiltonians, called {\it multipole models}, and study their properties.

Similar to the Kitaev honeycomb model \cite{Kitaev06}, multipole models arise in fragmented Hilbert spaces induced by local conserved quantities \cite{sala20,khemani20}. In spin models, the (smallest possible) local conserved quantities around site $\boldsymbol{r}$ are generally written as a product of $\nu$ spin operators $W_{\boldsymbol{r}}^{\nu}=\prod_{i=1}^{\nu}\sigma_{\boldsymbol{r}+\boldsymbol{a}_{i}}^{\mu_{i}}$, in which $\mu_{i}=x,y,z$ labels the spin components and $\boldsymbol{a}_{i}$ are lattice vectors. For $\nu\geq 3$, these quantities usually form closed loops and become $1$-form symmetries \cite{Gaiotto15,Mcgreevy23}; these are the cases of Kitaev-type spin models, including the original model \cite{Kitaev06}, the chiral spin liquid models \cite{yao2007,Fu20191}, and the Kitaev ladders \cite{DeGottardi11,Pedrocchi12,Wu2012}. Complementing this series of $\nu$, our multipole models form the simplest cases with $\nu=2$ local conserved quantites; geometrically, these quantities $\tau_{ij}=\sigma^{z}_{i}\sigma^{z}_{j}$ form {\it dimers}, each $\tau_{ij}$ takes values $\pm 1$. For dimers with $\tau=-1$, the allowed configurations are naturally {\it dipoles}; for dimers with $\tau=1$, they are two identical spins/charges, we name them {\it charge-pairs}. Each dimer has local invariant Hilbert subspace of dimension $2$, so dipoles or charge-pairs can be described by a spin-$\frac{1}{2}$ DOF \cite{Hotta10,Naka10}, representing the dipole moment for $\tau=-1$ sector and the sign of the charge-pair for $\tau=1$ sector; we call it the {\it first-order spin} while the original spin $\boldsymbol{\sigma}$ is taken as the zeroth-order. The original spin pair $\{\boldsymbol{\sigma}_{i},\boldsymbol{\sigma}_{j}\}$ is thus described by static $\mathbb{Z}_{2}$ variable $\tau_{ij}$ and first-order spin $\boldsymbol{S}_{ij}$.

In this work we study multipole models on ladder geometry. Although they can be introduced on almost all lattices, multipole models on ladders enjoy simple solubility. Our starting point is the {\it dipole models} with quadratic spin couplings, which map exactly to different types of transverse-field Ising model of first-order spin $\boldsymbol{S}$, coupled with static variables $\tau$. Adding certain plaquette terms with quartic spin couplings gives rise to general transverse-field XYZ model of first-order spin. The dipole models exhibit rich variety of phases characterized by the values of $\tau$ and the spin phases of $\boldsymbol{S}$. Generalizations to higher-order multipole models are achieved by introducing more local dimer conserved quantities. In general, multipole phases inherit physical properties of traditional spin phases of $\boldsymbol{S}$.

By revealing the relation between dimer local conserved quantities and multipole phases, our results shed light on a class of novel phases formed by composite objects. What makes them more interesting is that they are potentially realizable in experiments. As a fast developing field, Rydberg atom arrays are ideal platforms to simulate different types of spin systems \cite{Saffman10,bernien2017,zeiher2016}, including Ising and XY Hamiltonians \cite{Sylvain18,browaeys2020} as well as the XXZ Hamiltonian\cite{Nguyen18}. More complex spin interaction can be achieved with laser-assisted processes \cite{Glaetzle15,Chen2024,Yang2022}. For ladder geometries specifically, the Rydberg simulation has been considered theoretically \cite{Tsitsishvili22,fromholz22,Eck23,Liao25,Madhumita23} and realized experimentally \cite{zhang2025}. In this work we propose a simple design for simulating the dipole models with quadratic spin couplings using electric-field controlled Rydberg ladder \cite{Carroll04,Gonifmmode16,Comparat2010,Eiles17,Jahangiri2020}; such results open doors to artificially engineering dipole models in laboratories.

\section{Dipole models}

\begin{figure}
\includegraphics[width=0.5\textwidth]{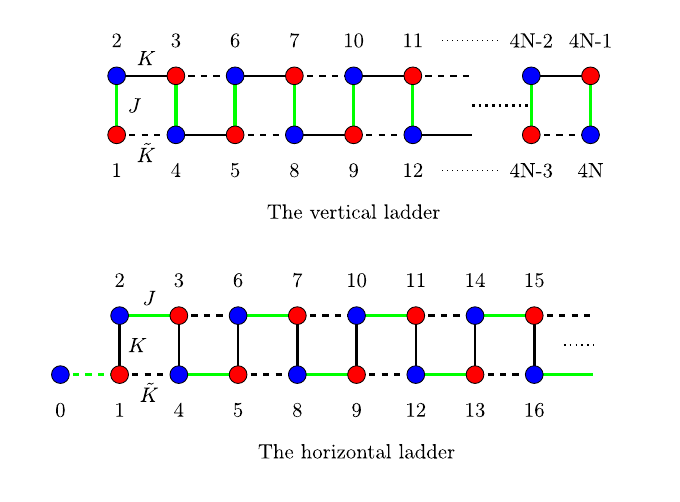}
\caption{Two ladder geometries for dipole models with quadratic spin couplings. The numbering of the sites marks the zig-zag string, along which the Jordan-Wigner transformations are performed. The $J$-bonds are represented by green solid lines while $K$-bonds are black lines (solid lines for coupling constant $K$ and dashed for $\tilde{K}$).}
\label{figladders}
\end{figure}

We start by considering dipole models on ladder geometry. The lattice bonds are divided into $J$-type with spin $XYZ$ coupling and $K$-type with Ising coupling. Dimer conserved quantities exist on every $J$-bond, the collection of which form a dimer covering of the ladder. The first possible geometry is the vertical ladder with all the vertical bonds as $J$-bonds and the rest as $K$-bonds. As shown in Fig. \ref{figladders}, the sites are linked together by a (virtual) zigzag string with numbers, every unit cell (vertical bonds) contains two spins. We consider a finite ladder with $4N$ individual spins, the Hamiltonian reads
\begin{eqnarray}
\label{modelone}
    \begin{aligned}
        \mathcal{H}_{v}^{(d)}=&\sum_{n=1}^{2N}\sum_{\alpha=x,y,z}J_{\alpha}\sigma_{2n-1}^{\alpha}\sigma_{2n}^{\alpha}\\&+\sum_{n=1}^{2N-1}\left(K\sigma_{2n}^{z}\sigma_{2n+1}^{z}+\tilde{K}\sigma_{2n-1}^{z}\sigma_{2n+2}^{z}\right).
    \end{aligned}
\end{eqnarray}
The model can be solved by Jordan-Wigner transformation in a rotated basis \cite{Fu2022}. Introducing two Majorana fermions $\eta^{\alpha}$ and $\eta^{\beta}$ for every site, we define
\begin{equation}
\label{JWtrans}
\sigma_{i}^{x}=-i\eta_{i}^{\alpha}\eta_{i}^{\beta},\qquad \sigma_{i}^{y}=\eta_{i}^{\alpha}\prod_{j=s}^{i-1}(i\eta_{j}^{\alpha}\eta_{j}^{\beta})
\end{equation}
with the starting point labeled by $s$; here we take $s=1$. The $\mathbb{Z}_{2}$ conserved quantities on $J$-bonds are
$\tau_{n}=\sigma_{2n-1}^{z}\sigma_{2n}^{z}=-i\eta_{2n-1}^{\alpha}\eta_{2n}^{\beta}$. With these the Hamiltonian is given by \cite{SM}
\begin{eqnarray}
\label{spinladderH}
    \begin{aligned}
        \mathcal{H}_{v}^{(d)}=&\sum_{n=1}^{2N}\left[(J_{y}-J_{x}\tau_{n})i\eta_{2n-1}^{\beta}\eta_{2n}^{\alpha}+J_{z}\tau_{n}\right]\\&+\sum_{n=1}^{2N-1}\left(K+\tilde{K}\tau_{n}\tau_{n+1}\right)(-i)\eta_{2n}^{\alpha}\eta_{2n+1}^{\beta},
    \end{aligned}
\end{eqnarray}
which is a free fermion model coupled to $\tau$ variables. We pair up every $J$-bond with two sites $(2n-1,2n)$ and label it with number $n$, then introduce the first-order spin $\boldsymbol{S}_{n}$ for every $J$-bond by a new JW transformation: $\eta_{2n}^{\alpha}=S_{n}^{y}\prod_{j=1}^{n-1}(-S_{j}^{x})$, $\eta_{2n-1}^{\beta}=S_{n}^{z}\prod_{j=1}^{n-1}(-S_{j}^{x})$. The Hamiltonian \eqref{spinladderH} becomes
\begin{eqnarray}
\label{transverseising}
    \begin{aligned}
        \mathcal{H}_{v}^{(d)}=&\sum_{n=1}^{2N}\left[(J_{y}-J_{x}\tau_{n})S_{n}^{x}+J_{z}\tau_{n}\right]\\&+\sum_{n=1}^{2N-1}(K+\tilde{K}\tau_{n}\tau_{n+1})S_{n}^{z}S_{n+1}^{z},
    \end{aligned}
\end{eqnarray}
which is identical to the transverse-field Ising model \cite{Suzuki13} coupled with static $\tau$ variables.
The first-order spin $\boldsymbol{S}_{n}$ describes the dipole and charge-pair DOF within each local invariant sector, they are related to the zeroth-order spin $\boldsymbol{\sigma}_{i}$ by \cite{SM}
\begin{equation}
\label{relationvertical}
    S_{n}^{x}=\sigma_{2n-1}^{y}\sigma_{2n}^{y},\qquad S_{n}^{z}=\prod_{j=1}^{2n-1}\sigma_{j}^{z}.
\end{equation}
Eq. \eqref{relationvertical} tells that the flipping of $S_{n}^{z}$ is equivalent to the flipping of $\sigma_{2n-1}^{z}$ and $\sigma_{2n}^{z}$ together, namely the flipping of dipole/charge-pair within each local sector. To capture such flip, we define a global {\it dipole parity}, $\mathcal{I}=\prod_{n=1}^{2N}S_{n}^{z}=\prod_{n=1}^{2N}\prod_{j=1}^{2n-1}\sigma_{j}^{z}$,
which satisfies $\{S_{n}^{x},\mathcal{I}\}=0$.

\begin{figure}
\includegraphics[width=0.35\textwidth]{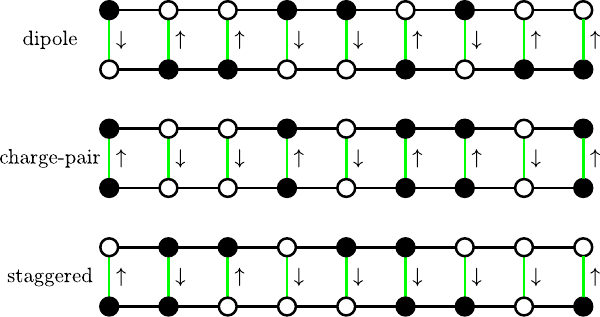}
\includegraphics[width=0.45\textwidth]{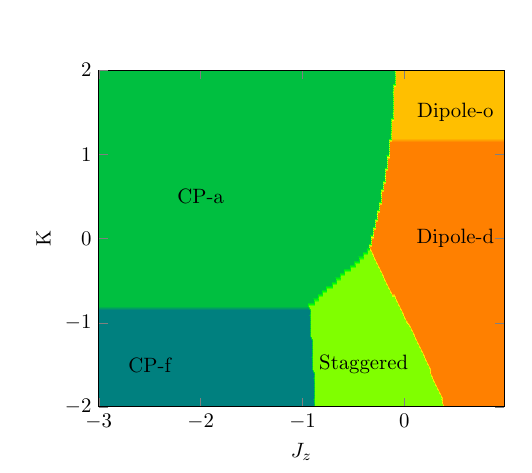}
\caption{Phases of the vertical ladder. Upper: illustration of different phases with arrows representing the first-order spin $S^{z}$, filled (empty) circles representing up (down) spins. Lower: phase diagram of the vertical ladder (quadratic spin coupling) with $\tilde{K}=0.8$, $J_{x}=J_{y}=1$ obtained from the fermionic ground-state energy in \eqref{spinladderH}.}
\label{figvertical}
\end{figure}

Depending on the ground-state configuration, the system can reside in the {\it dipole phase}, where all $\tau_{n}$ equal $-1$, the {\it charge-pair phase} with all $\tau_{n}=1$, or the {\it staggered phase}, in which $\tau_{n}$ distribution breaks translational symmetry. The phases are further classified by the configuration of the first-order spin $\boldsymbol{S}_{n}$, as shown in Fig. \ref{figvertical}. To obtain the phase diagram we focus on a specific choice of parameters $J_{x}=J_{y}=1$ and $\tilde{K}=0.8$. For these parameters, the transverse term in \eqref{transverseising} vanishes for the charge-pair phase and it effectively becomes a classical Ising model, so we have the ``charge-pair-antiferromagnetic (CP-a)" phase and the ``charge-pair-ferromagnetic (CP-f)" phase. In the dipole phase, we always have zero net-magnetization $\langle \sigma^{z}\rangle=0$, but the system can still be ordered for dipole moment $\boldsymbol{S}$, indicated by the correlation function $\langle S_{n}^{z}S_{n+m}^{z}\rangle=(\prod_{j=1}^{m-1}\tau_{n+j})\langle \sigma_{2n}^{z}\sigma_{2m+2n-1}^{z}\rangle$. Due to the non-vanishing transverse term in \eqref{transverseising}, there is a phase transition between the ordered and disordered phases originated from the transverse-field-Ising model, which has topological nature. These are called the ``dipole-disordered (Dipole-d)" phase and the ``dipole-ordered (Dipole-o)" phase. The phase diagram is given in Fig. \ref{figvertical}. {\it Weak zero mode} exists in Dipole-o phase \cite{Fu2022}, corresponding to the Ising zero mode \cite{Greiter14,Niu12,Fu2021} for the dipole moment (first-order spin) $\boldsymbol{S}_{n}$; it must be bosonic in terms of $\boldsymbol{\sigma}$. 

More general dipole (and charge-pair) phases arise when adding certain plaquette terms with quartic spin coupling to the Hamiltonian \eqref{modelone}. The relation \eqref{relationvertical} tells that XX and YY couplings of the first-order spin translate to quartic terms of $\boldsymbol{\sigma}$, namely $S_{n}^{x}S_{n+1}^{x}=\sigma_{2n-1}^{y}\sigma_{2n}^{y}\sigma_{2n+1}^{y}\sigma_{2n+2}^{y}$ and $S_{n}^{y}S_{n+1}^{y}=\sigma_{2n-1}^{y}\sigma_{2n}^{x}\sigma_{2n+1}^{x}\sigma_{2n+2}^{y}$. The extended Hamiltonian \eqref{modelone} with these plaquette terms can support phases of the general XYZ model in transverse field for the first-order spin $\boldsymbol{S}$ \cite{SM}. Low-energy excitations are usually given by traditional spin excitations of first-order spin $\boldsymbol{S}$ depending on the specific ground-state sector.

Another possible geometry is the horizontal ladder shown in Fig. \ref{figladders}. For definition purpose, the ladder is taken as half-infinite and has an extra spin site (site 0). The Hamiltonian is 
\begin{eqnarray}
\label{horizontalH}
    \begin{aligned}
        \mathcal{H}_{h}^{(d)}=&\sum_{n=0}\sum_{\alpha}J_{\alpha}\sigma_{2n}^{\alpha}\sigma_{2n+1}^{\alpha}\\&+\sum_{n=1}\left[K\sigma_{2n-1}^{z}\sigma_{2n}^{z}+\tilde{K}\sigma_{2n-1}^{z}\sigma_{2n+2}^{z}\right].
    \end{aligned}
\end{eqnarray}
Introducing Majorana fermions $\eta^{\alpha}$, $\eta^{\beta}$ on every site, the JW transformation (\ref{JWtrans}) with starting point $s=0$ brings \eqref{horizontalH} into a fermionic Hamiltonian. The $\mathbb{Z}_{2}$ conserved quantities on every horizontal $J-$bond are $\tau_{n}=\sigma_{2n}^{z}\sigma_{2n+1}^{z}=-i\eta_{2n}^{\alpha}\eta_{2n+1}^{\beta}$. With these the Hamiltonian \eqref{horizontalH} is brought into an interacting fermionic model coupled to static variables $\tau$ \cite{SM}. We then pair up all the $J-$bonds with sites $(2n,2n+1)$ and label them with $n=0,1,2,\cdots$. The first-order spin $\boldsymbol{S}_{n}$ is defined for each pair ($J-$bond) by a new JW transformation $\eta_{2n+1}^{\alpha}=S_{n}^{y}\prod_{j=0}^{n-1}(-S_{j}^{x})$, $\eta_{2n}^{\beta}=S_{n}^{z}\prod_{j=0}^{n-1}(-S_{j}^{x})$. The Hamiltonian \eqref{horizontalH} thus becomes
\begin{eqnarray}
\label{HtwoS}
    \begin{aligned}
        \mathcal{H}_{h}^{(d)}=&\sum_{n=0}\left[(J_{y}-J_{x}\tau_{n})S_{n}^{x}+J_{z}\tau_{n}\right]\\&+\sum_{n=1}\left[KS_{n-1}^{z}S_{n}^{z}+\tilde{K}\tau_{n}S_{n-1}^{z}S_{n+1}^{z}\right],
    \end{aligned}
\end{eqnarray}
which involves Ising interaction of second-nearest neighbors. It is the axial next-nearest-neighbor Ising (ANNNI) model in transverse field \cite{Selke88,Fisher80,Hornreich79,Sen89,Suzuki13} coupled to static $\mathbb{Z}_{2}$ variables. The first-order spin $\boldsymbol{S}$ is related to the zeroth-order spin $\boldsymbol{\sigma}$ by $S_{n}^{z}=\prod_{j=0}^{2n}\sigma_{j}^{z}$, and $S_{n}^{x}=\sigma_{2n}^{y}\sigma_{2n+1}^{y}$. 

\begin{figure}
\includegraphics[width=0.3\textwidth]{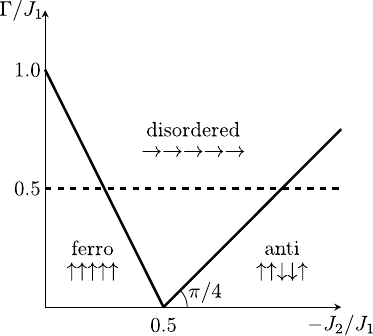}
\includegraphics[width=0.45\textwidth]{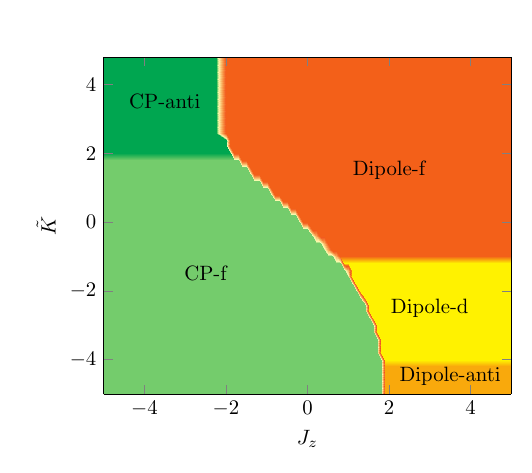}
\caption{Phases of the horizontal ladder. Upper: {\it schematic} phase diagram of the ANNNI chain which we adopt in this work. Lower: phase diagram of the horizontal ladder with $K=-4$, $J_{x}=J_{y}=-1$ (corresponding to the dashed line in the upper figure).}
\label{fighorizontal}
\end{figure}

The Hamiltonian \eqref{HtwoS} can be compared with the standard form of the ANNNI model, $\mathcal{H}^{I}=-\sum_{i}J_{1}S_{i}^{z}S_{i+1}^{z}-\sum_{i}J_{2}S_{i}^{z}S_{i+2}^{z}-\sum_{i}\Gamma S_{i}^{x}$, correspondingly $J_{1}=-K$, $J_{2}=-\tilde{K}\tau_{n}$ and $\Gamma=-(J_{y}-J_{x}\tau_{n})$. The phases of the ANNNI model have been studied \cite{Arizmendi91,Sen91,Sen92,Suzuki13}, here we take a schematic (simplified) phase diagram shown in Fig. \ref{fighorizontal}. Compared with the Ising model, the ANNNI model has a special ordered phase, called ``anti-phase", in which spin order reads two ups and two downs. Notably the phase boundaries meet at $-\frac{J_{2}}{J_{1}}=0.5$, $\Gamma=0$, which is the Majumdar-Ghosh point \cite{Magh1,Magh2}. To study the phase diagram of the horizontal ladder \eqref{HtwoS}, we take $J_{x}=J_{y}=-1$ and $K=-4$, corresponding to $\Gamma/J_{1}=0.5$ for the dipole phase and $\Gamma/J_{1}=0$ for the charge-pair phase. The phase diagram from diagonalizing a system of 16 spins (with periodic boundary condition) is shown in Fig. \ref{fighorizontal}.

\section{Quadrupole models and beyond}

\begin{figure}
\includegraphics[width=0.47\textwidth]{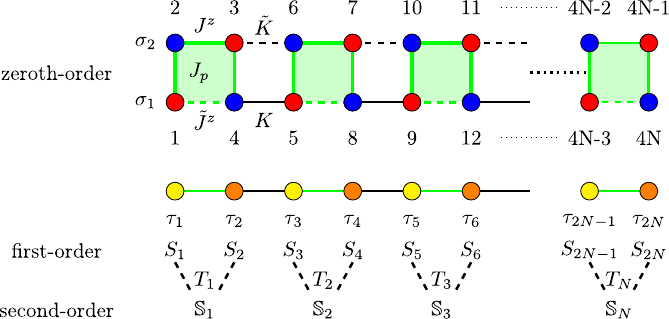}
\caption{The quadrupole ladder with zeroth-order spin $\sigma$. The plaquette terms are represented by the green squares; green lines mark the ZZ coupling that commute with the Hamiltonian. Introducing first-order spin $S$, the model becomes a dipole chain which is further brought down to transverse-field Ising model with second-order spin $\mathbb{S}$.}
\label{figquadrupole}
\end{figure}

The dipole phases protected by $\nu=2$ local conserved quantities can be generalized to models with quadrupole phases. We consider the following quadrupole model defined on the ladder geometry shown in Fig. \ref{figquadrupole},
\begin{eqnarray}
\label{higherorderH}
\begin{aligned}
\mathcal{H}^{(q)}=&\sum_{n=1}^{N}(J_{p}^{x}\prod_{j=4n-3}^{4n}\sigma_{j}^{x}+J_{p}^{y}\prod_{j=4n-3}^{4n}\sigma_{j}^{y})\\&+\sum_{n=1}^{N}(J^{z}\sigma_{4n-2}^{z}\sigma_{4n-1}^{z}+\tilde{J}^{z}\sigma_{4n-3}^{z}\sigma_{4n}^{z})\\&+\sum_{n=1}^{N-1}(K\sigma_{4n}^{z}\sigma_{4n+1}^{z}+\tilde{K}\sigma_{4n-1}^{z}\sigma_{4n+2}^{z}).
\end{aligned}
\end{eqnarray}
The Hamiltonian has plaquette terms quartic in spin on every second square of the ladder as well as the Ising terms appearing in \eqref{modelone} and \eqref{horizontalH}. This model has dimer conserved quantities not only on the vertical bonds $(2n-1,2n)$ but also on half of the horizontal bonds $(4n-2,4n-1)$ and $(4n-3,4n)$ for $n=1\cdots N$; together they form square plaquettes. Using the relation \eqref{relationvertical} and the conserved quantities on vertical bonds $\tau_{n}=\sigma_{2n-1}^{z}\sigma_{2n}^{z}$, the Hamiltonian (\ref{higherorderH}) can be written in terms of first-order spin $\boldsymbol{S}$, $\mathcal{H}^{(q)}=\sum_{n=1}^{N}(J_{p}^{x}\tau_{2n-1}\tau_{2n}+J_{p}^{y})S_{2n-1}^{x}S_{2n}^{x}+\sum_{n=1}^{N}(J^{z}+\tilde{J}^{z}\tau_{2n-1}\tau_{2n})S_{2n-1}^{z}S_{2n}^{z}+\sum_{n=1}^{N-1}(K+\tilde{K}\tau_{2n}\tau_{2n+1})S_{2n}^{z}S_{2n+1}^{z}$, which is a dipole model \cite{Fu2022}. It has dimer conserved quantities of first-order spin $T_{n}=S_{2n-1}^{z}S_{2n}^{z}$, with $[T_{n},\mathcal{H}^{(q)}]=0$. These bilinears translate back to the zeroth-order spin $T_{n}=\sigma_{4n-2}^{z}\sigma_{4n-1}^{z}$, which correspond to the additional conserved quantities on the horizontal bonds in \eqref{higherorderH}. 

The quadrupole model \eqref{higherorderH} can be further brought down to a transverse-field Ising model. We first introduce the {\it second-order} spin $\boldsymbol{\mathbb{S}}$ by \eqref{relationvertical}, namely $\mathbb{S}_{n}^{x}=S_{2n-1}^{y}S_{2n}^{y}$ and $\mathbb{S}_{n}^{z}=\prod_{j=1}^{2n-1}S_{j}^{z}$, with which the Hamiltonian \eqref{higherorderH} becomes
\begin{eqnarray}
\label{higherorderIsing}
    \begin{aligned}
        \mathcal{H}^{(q)}=&\sum_{n=1}^{N}-(J_{p}^{x}\tau_{2n-1}\tau_{2n}+J_{p}^{y})T_{n}\mathbb{S}_{n}^{x}\\
        &+\sum_{n=1}^{N}(J^{z}+\tilde{J}^{z}\tau_{2n-1}\tau_{2n})T_{n}\\
        &+\sum_{n=1}^{N-1}(K+\tilde{K}\tau_{2n}\tau_{2n+1})\mathbb{S}_{n}^{z}\mathbb{S}_{n+1}^{z}.
    \end{aligned}
\end{eqnarray}
The quadrupole model of zeroth-order spin $\boldsymbol{\sigma}$ is transformed into a dipole model of first-order spin $\boldsymbol{S}$, then into transverse-field Ising model of second-order spin $\mathbb{S}$, this process can be denoted by $\mathcal{H}^{(q)}(\sigma)\rightarrow \mathcal{H}^{(d)}(S)\rightarrow \mathcal{H}^{I}(\mathbb{S})$.

Phases of the quadrupole model \eqref{higherorderH} are first classified by the static $\mathbb{Z}_{2}$ variables $\tau$ and $T$, for example, $\tau_{2n-1}=\tau_{2n}=T_{n}=-1$ is ``Quadrupole". For plaquette $n$ in the original model \eqref{higherorderH}, its zeroth-order spin configuration is partially fixed by two $\mathbb{Z}_{2}$ quantities $\tau_{2n-1}$ and $\tau_{2n}$ together with another $\mathbb{Z}_{2}$ quantity $T_{n}$; the local space is divided into $2^{3}=8$ independent sectors, the remaining dimension $16/8=2$ is the second-order spin $\mathbb{S}_{n}$. Similar to the dipole models, phases of the zeroth-order spin $\boldsymbol{\sigma}$ in the original model \eqref{higherorderH} are further determined by transverse-Ising phases of the second-order spin $\mathbb{S}$ in \eqref{higherorderIsing}. Details are found in the Supplemental Materials \cite{SM}. More general phases of $\mathbb{S}$ are obtainable by including higher-order terms of $\boldsymbol{\sigma}$ in \eqref{higherorderH}.

Further down the road, generalization to octopole models can be done by coupling to another layer of ladder, with the plaquettes becoming cubes. The Hamiltonian, containing cubic, plaquette and bond terms, has $\mathbb{Z}_{2}$ conserved quantities on every edge of half of the cubes \cite{SM}. In principle, such generalization goes on to higher-order, and $n$-th order spin describes dipole/charge-pair formed by the $(n-1)$-th order spin. Higher-order models generally have more complex phase structure.

\section{Rydberg simulation of the dipole model}

\begin{figure}
\includegraphics[width=0.43\textwidth]{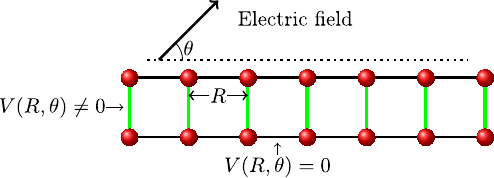}
\caption{The simulation of the dipole model on vertical ladder using electric field controlled Rydberg atom array. The direction of the electric field and the distance between neighboring atoms are tuned such that the exchange coupling $V(R,\theta)$ vanishes on horizontal bonds.}
\label{figRyd}
\end{figure}

Rydberg atom array is the ideal platform to simulate the dipole models studied in this work. Here we propose a simple design aiming at the dipole model on vertical ladder \eqref{modelone}. In particular we use electric-field controlled Rydberg states to produce the bond-dependent interaction \cite{Carroll04,Gonifmmode16,Comparat2010,Schwettmann06,Eiles17}. The atoms are trapped in a ladder geometry with optical tweezers. A small electric field is applied so there is no significant mixing between states, and its direction is at angle $\theta$ with the horizontal axis of the ladder, as shown in Fig. \ref{figRyd}. The spin states are encoded in two Stark-shifted Rydberg $S$ states with different principle quantum numbers, namely $|\uparrow\rangle\rightarrow |\widetilde{n_{\uparrow}S}\rangle$ and $|\downarrow\rangle\rightarrow |\widetilde{n_{\downarrow}S}\rangle$. Without electric field, the interactions between nearest neighboring atoms are purely van der Waals (vdW), which gives rise to an effective spin XXZ model \cite{Whitlock17,Bijnen13}. 

Inside the electric field, the atoms directly couple by dipole-dipole interaction, which is first order as opposed to the second-order vdW interaction. Take two atoms separated by $R$, the matrix element of the effective Hamiltonian causing the spin exchange reads $\langle\uparrow\downarrow|\mathcal{H}|\downarrow\uparrow\rangle=V(R,\theta)$, and $V(R,\theta)=\frac{C_{6}}{R^{6}}+\frac{C_{3}}{R^{3}}(1-3\cos^{2}\theta)$ \cite{Gonifmmode16,Jiao2022,Jahangiri2020} with $C_{6}$ and $C_{3}$ being couplings of vdW and dipole-dipole interaction. And the effective XY coupling strength is $J_{xy}=V(R,\theta)/2$ on all bonds of the lattice \cite{Whitlock17,Bijnen13} (more details are presented in the Supplemental Materials \cite{SM}). As the angle between the electric field and the horizontal lattice bonds is different from that for the vertical bonds, the effective XY couplings are different and can be further tuned by the distances between the atoms. Our goal is to tune the distance $R$ and angle $\theta$ such that $V(R,\theta)=0$ for the horizontal bonds. When this is achieved, the coupling on the horizontal bonds will be Ising while the coupling on the vertical bonds remains XXZ. Strictly speaking there is another term generated ($\sum_{i}h\sigma_{i}$) which corresponds to an effective external field. Such term originates from the asymmetry between the two spin states \cite{SM} and can in principle be compensated by applying a tunable microwave field \cite{Whitlock17}. we thus obtained the dipole model on vertical ladder \eqref{modelone} (see Fig. \ref{figRyd}). In this preliminary discussion, we have neglected the atomic interaction beyond nearest neighbors. Although decaying as fast as $R^{3}$ and $R^{6}$, these interactions certainly exist and their effects on the dipole phases should be addressed in future studies.

\section{Discussion}

We have shown the relation between multipole phases and local dimer conserved quantities through dipole and quadrupole models on ladder geometry, our results pave the way towards engineering traditional spin phases with multipole moments. Although we focus on spin description throughout this paper, the dual fermionic version of the models can be obtained by standard Jordan-Wigner transformation \cite{Jordan1928,Lieb61,Feng07,Fu2022,SM}. Besides it is also interesting to study the multipole phases on higher-dimensional lattices; similar multipole phases are expected to exist although JW transformation does not apply easily. 

Besides the Rydberg atom arrays, the dipole model can be realized in other platforms, such as ladder superconducting devices \cite{Wang2025}. Except for artificial simulations, it's also appealing to look for such models in strongly correlated materials. If such materials exist, the phases of the dipole models can be distinguished experimentally in neutron scattering \cite{Marshall68,girvinbook} and magnetic Raman measurements \cite{Fu17,Ho10}. For neutron scattering we compute the zero-temperature dynamical structure factor (DSF) \cite{Derzhko97,Caux03,James09,chakra06,Pereira07,Haga02}. For the vertical ladder, the zero-temperature DSF can be written as \cite{SM} $D^{zz}(q_{x},q_{y},\omega)=(1\pm\cos q_{y})D_{0}^{zz}(q_{x},\omega)$, in which $+$ sign is for the charge-pair phase, $-$ sign is for the dipole phase and $D_{0}^{zz}(q_{x},\omega)$ is the DSF for the Ising chain \cite{Derzhko97,chakra06,Young97} (see Supplemental Materials \cite{SM} for details). Notably the dipole-model DSF {\it vanishes} at different series of momenta on the $q_{x}-q_{y}$ plane for different phases, these can be used to pin-point the phases experimentally.

\section*{Acknowledgements}

The author would like to thank Hoi Chun Po for valuable discussion. This work is supported by National Key R \& D Program of China (Grants No.2021YFA1401500).


\bibliography{refladder}

\clearpage
\widetext
\begin{center}
{\Large \textbf{Supplemental Materials}}
\end{center}
\setcounter{equation}{0}
\setcounter{figure}{0}
\setcounter{table}{0}
\setcounter{page}{1}
\makeatletter
\renewcommand{\theequation}{S\arabic{equation}}
\renewcommand{\thefigure}{S\arabic{figure}}

\section*{Interacting fermionic version of the dipole models}

The dipole models on the ladder geometry with Hamiltonians 
\begin{equation}
    \mathcal{H}_{\sigma}^{(d)}=\sum_{J\text{-bonds}}\left(J_{x}\sigma_{i}^{x}\sigma_{j}^{x}+J_{y}\sigma_{i}^{y}\sigma_{j}^{y}+J_{z}\sigma_{i}^{z}\sigma_{j}^{z}\right)+\sum_{K\text{-bonds}}K_{ij}\sigma_{i}^{z}\sigma_{j}^{z}.
\label{HsigmaS}
\end{equation}
are equivalent to fermionic models with identical spectrum. We first introduce a zigzag string linking all the sites together, the geometries of the vertical ladder and horizontal ladder can then by represented by Fig. \ref{figSladder}. A standard Jordan-Wigner (JW) transformation is performed along the string by introducing a fermion $c_{i}$ on every site and relating it to the spin by $\sigma_{i}^{+}=c_{i}^{\dagger}(-1)^{\sum_{j=1}^{i-1}c_{j}^{\dagger}c_{j}}$ and $\sigma_{i}^{z}=2c_{i}^{\dagger}c_{i}-1$. The resulting fermionic Hamiltonian is particularly simple if the zigzag string connects the $J$-bonds and $K$-bonds intermediately (as for both the vertical ladder and the horizontal ladder), for which case the fermionic Hamiltonian corresponding to (\ref{HsigmaS}) reads (assuming the couplings are isotropic, $J_{x}=J_{y}=J_{xy}$)
\begin{equation}
\begin{aligned}
    \mathcal{H}_{c}^{(d)}=&\sum_{J\text{-bonds}}\bigg[2J_{xy}(c_{i}^{\dagger}c_{j}+c_{j}^{\dagger}c_{i})+J_{z}(2c_{i}^{\dagger}c_{i}-1)(2c_{j}^{\dagger}c_{j}-1)\bigg]\\&+\sum_{K\text{-bonds}}K_{ij}(2c_{i}^{\dagger}c_{i}-1)(2c_{j}^{\dagger}c_{j}-1).
\end{aligned}
\label{Hc}
\end{equation}
This is a general fermionic model with interaction on all bonds and fermion-hopping exists only on the dimers, the superscript $(d)$ means {\it dipole}.

\section*{Details of the solution of the models}

\subsection{The vertical ladder}

\begin{figure}
\includegraphics[width=\textwidth]{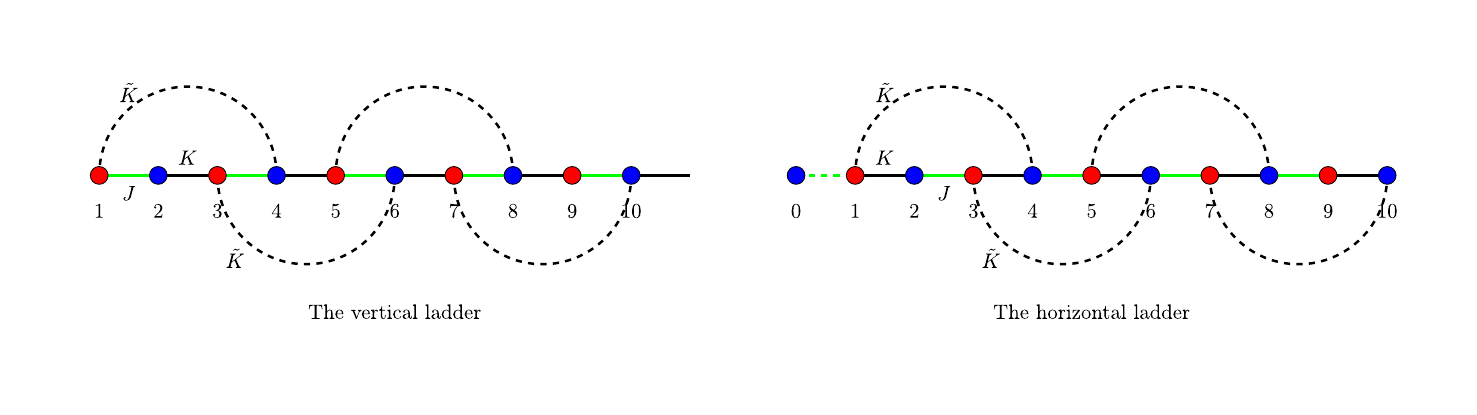}
\caption{Two ladder geometries considered in this work presented in a different way. The numbering of the sites marks the zig-zag string, along which the JW transformations are performed. The $J$-bonds (XYZ coupling) are represented by green solid lines, $K$-bonds (ZZ coupling) are black lines, both solid and dashed lines are used to mark the two coupling constants $K$ and $\tilde{K}$.}
\label{figSladder}
\end{figure}

For the vertical ladder, the vertical bonds have spin XYZ interaction and the horizontal bonds have spin ZZ coupling. Every unit cell (vertical bonds) cotains two spins; we consider $2N$ original unit cells and $4N$ individual spins. Fig. \ref{figSladder} gives another way to represent the ladder geometry. The Hamiltonian of the spin ladder is given by 
\begin{equation}
\label{SmodeloneS}
\mathcal{H}_{v}^{(d)}=\sum_{n=1}^{2N}\left(J_{x}\sigma_{2n-1}^{x}\sigma_{2n}^{x}+J_{y}\sigma_{2n-1}^{y}\sigma_{2n}^{y}+J_{z}\sigma_{2n-1}^{z}\sigma_{2n}^{z}\right)+\sum_{n=1}^{2N-1}\left(K\sigma_{2n}^{z}\sigma_{2n+1}^{z}+\tilde{K}\sigma_{2n-1}^{z}\sigma_{2n+2}^{z}\right).
\end{equation}
The model has $\mathbb{Z}_{2}$ conserved quantities $\sigma_{2n-1}^{z}\sigma_{2n}^{z}$ for every vertical bond, it can be solved by the JW transformation in a rotated basis. In particular we introduce two Majorana fermions $\eta^{\alpha}$ and $\eta^{\beta}$ for every site and the JW transformation is given by, defining the starting point as $s$, 
\begin{equation}
\label{JWsupplemental}
\sigma_{i}^{x}=-i\eta_{i}^{\alpha}\eta_{i}^{\beta},\qquad \sigma_{i}^{y}=\eta_{i}^{\alpha}\prod_{j=s}^{i-1}(i\eta_{j}^{\alpha}\eta_{j}^{\beta}),\qquad \sigma_{i}^{z}=\eta_{i}^{\beta}\prod_{j=s}^{i-1}(i\eta_{j}^{\alpha}\eta_{j}^{\beta}).
\end{equation}
Applying the JW transformation with $s=1$, the Hamiltonian can be written as 
\begin{eqnarray}
\begin{aligned}
\mathcal{H}_{v}^{(d)}=&\sum_{n=1}^{2N}\left[J_{x}(-)(\eta_{2n-1}^{\alpha}\eta_{2n}^{\beta})\eta_{2n-1}^{\beta}\eta_{2n}^{\alpha}+J_{y}(-i)\eta_{2n}^{\alpha}\eta_{2n-1}^{\beta}+J_{z}i(\eta_{2n}^{\beta}\eta_{2n-1}^{\alpha})\right]+\\&\sum_{n=1}^{2N-1}\left[K(-i)\eta_{2n}^{\alpha}\eta_{2n+1}^{\beta}+\tilde{K}(-i)\eta_{2n+2}^{\beta}\eta_{2n-1}^{\alpha}\eta_{2n}^{\alpha}\eta_{2n}^{\beta}\eta_{2n+1}^{\alpha}\eta_{2n+1}^{\beta}\right],
\end{aligned}
\end{eqnarray}
and the conserved quantities are $\tau_{n}=\sigma_{2n-1}^{z}\sigma_{2n}^{z}=-i\eta_{2n-1}^{\alpha}\eta_{2n}^{\beta}$.

For each sector with definite $\tau_{n}$, the Hamiltonian is identical to the transverse-field Ising model. By pairing up sites $(2n-1,2n)$ into a single site $n$ (namely $(2n-1,2n)\rightarrow n$) and redefining the Majorana fermions as $\eta_{2n-1}^{\beta}\rightarrow \tilde{\eta}_{n}^{\beta}$ and $\eta_{2n}^{\alpha}\rightarrow\tilde{\eta}_{n}^{\alpha}$, the Hamiltonian then becomes
\begin{equation}
\label{fermionicH}
\mathcal{H}_{v}^{(d)}=\sum_{n=1}^{2N}\left[(J_{y}-J_{x}\tau_{n})i\tilde{\eta}_{n}^{\beta}\tilde{\eta}_{n}^{\alpha}+J_{z}\tau_{n}\right]+\sum_{n=1}^{2N-1}(K+\tilde{K}\tau_{n}\tau_{n+1})(-i)\tilde{\eta}_{n}^{\alpha}\tilde{\eta}_{n+1}^{\beta}.
\end{equation}
A new set of JW transformation can be performed by introducing the first-order spin $\boldsymbol{S}_{n}$,
\begin{equation}
\tilde{\eta}_{n}^{\alpha}=S_{n}^{y}\prod_{j=1}^{n-1}(-S_{j}^{x}),\qquad \tilde{\eta}_{n}^{\beta}=S_{n}^{z}\prod_{j=1}^{n-1}(-S_{j}^{x}),\qquad i\tilde{\eta}_{n}^{\beta}\tilde{\eta}_{n}^{\alpha}=S_{n}^{x}.
\end{equation}
This leads to $(-i)\tilde{\eta}_{n}^{\alpha}\tilde{\eta}_{n+1}^{\beta}=S_{n}^{z}S_{n+1}^{z}$, using which we arrive at the Hamiltonian in the main text.

The first-order spin $\boldsymbol{S}_{n}$ can be related to the zeroth-order spin $\boldsymbol{\sigma}$, firstly 
\begin{equation}
\label{Srelationsigma}
S_{n}^{x}=i\eta_{2n-1}^{\beta}\eta_{2n}^{\alpha}=i[\sigma_{2n-1}^{z}\prod_{j=1}^{2n-2}(-\sigma_{j}^{x})]\sigma_{2n}^{y}\prod_{j=1}^{2n-1}(-\sigma_{j}^{x})=\sigma_{2n-1}^{y}\sigma_{2n}^{y}.
\end{equation}
On the other hand, we have $S_{n}^{z}=\eta_{2n-1}^{\beta}\prod_{j=1}^{n-1}(i\eta_{2j}^{\alpha}\eta_{2j-1}^{\beta})$, which is compared with
\begin{equation}
\sigma_{2n-1}^{z}=\eta_{2n-1}^{\beta}\prod_{j=1}^{2n-2}(i\eta_{j}^{\alpha}\eta_{j}^{\beta})=\eta_{2n-1}^{\beta}\prod_{j=1}^{n-1}(i\eta_{2j-1}^{\alpha}\eta_{2j}^{\beta})(i\eta_{2j-1}^{\beta}\eta_{2j}^{\alpha}),
\end{equation}
and yields
\begin{equation}
\sigma_{2n-1}^{z}=S_{n}^{z}\prod_{j=1}^{n-1}(-i\eta_{2j-1}^{\alpha}\eta_{2j}^{\beta})=S_{n}^{z}\prod_{j=1}^{n-1}\tau_{n}=S_{n}^{z}\prod_{j=1}^{2n-2}\sigma_{j}^{z}.
\end{equation}
Finally we get the Eq. (5) in the main text. 

In the main text, we define a global dipole parity, $\mathcal{I}=\prod_{n=1}^{2N}S_{n}^{z}=\prod_{n=1}^{2N}\prod_{j=1}^{2n-1}\sigma_{j}^{z}$, which satisfies $\{S_{n}^{x},\mathcal{I}\}=0$. Explicitly it can be written as $\mathcal{I}=\prod_{j=1}^{4N-1}\sigma_{j}^{2N-x_{j}}$, where $x_{j}$ is the {\it effective coordinates} of the sites. Along the zigzag string the effective length of the $J$-bonds is one while that of the $K$-bonds is zero.

In the main text, we obtain the phase diagram by exactly diagonalizing the fermionic Hamiltonian \eqref{fermionicH}. Here we provide the details of the derivation. Firstly we define complex fermion $f_{n}^{\dagger}=\frac{1}{2}(\tilde{\eta}_{n}^{\beta}+i\tilde{\eta}_{n}^{\alpha})$, such that $\tilde{\eta}_{n}^{\beta}=f_{n}+f_{n}^{\dagger}$, $\tilde{\eta}_{n}^{\alpha}=i(f_{n}-f_{n}^{\dagger})$. The Hamiltonian \eqref{fermionicH} can be written as
\begin{equation}
\mathcal{H}_{v}^{(d)}=\sum_{n=1}^{2N}\left[(J_{x}\tau_{n}-J_{y})(f_{n}+f_{n}^{\dagger})(f_{n}-f_{n}^{\dagger})+J_{z}\tau_{n}\right]+\sum_{n=1}^{2N-1}(K+\tilde{K}\tau_{n}\tau_{n+1})(f_{n}-f_{n}^{\dagger})(f_{n+1}+f_{n+1}^{\dagger}).
\end{equation}
Only phases which preserve the translational symmetry are considered. To this end, we pick the unit cell containing two $f$ fermions, and rename $f_{2n-1}\rightarrow f_{n}^{1}$, $f_{2n}\rightarrow f_{n}^{2}$, $\tau_{2n-1}\rightarrow \tau_{n}^{1}$, $\tau_{2n}\rightarrow \tau_{n}^{2}$. The Hamiltonian \eqref{fermionicH} then becomes 
\begin{eqnarray}
\begin{aligned}
\mathcal{H}_{v}^{(d)}=&\sum_{n=1}^{N}(J_{x}\tau_{n}^{1}-J_{y})(2f_{n}^{1\dagger}f_{n}^{1}-1)+(J_{x}\tau_{n}^{2}-J_{y})(2f_{n}^{2\dagger}f_{n}^{2}-1)+J_{z}(\tau_{n}^{1}+\tau_{n}^{2})\\&+(K+\tilde{K}\tau_{n}^{1}\tau_{n}^{2})\bigg(f_{n}^{1}f_{n}^{2}-f_{n}^{1\dagger}f_{n}^{2}+f_{n}^{1}f_{n}^{2\dagger}-f_{n}^{1\dagger}f_{n}^{2\dagger}\bigg)\\
&+\sum_{n=1}^{N-1}(K+\tilde{K}\tau_{n}^{2}\tau_{n+1}^{1})\bigg(f_{n}^{2}f_{n+1}^{1}-f_{n}^{2\dagger}f_{n+1}^{1}+f_{n}^{2}f_{n+1}^{1\dagger}-f_{n}^{2\dagger}f_{n+1}^{1\dagger}\bigg).
\end{aligned}
\end{eqnarray}
Due to the assumed translational symmetry, a Fourier transform can be performed by defining $f_{n}^{\dagger}=\frac{1}{\sqrt{N}}\sum_{k}f_{k}^{\dagger}e^{ikn}$, the range of $k$ is taken to be $(-\pi,\pi)$. For every unit cell, the values of $\tau^{1}$ and $\tau^{2}$ are fixed and independent of the position, so we have 
\begin{eqnarray}
\label{fIsing}
\begin{aligned}
\mathcal{H}_{v}^{(d)}=\sum_{k}&(J_{x}\tau^{1}-J_{y})(2f_{k}^{1\dagger}f_{k}^{1}-1)+(J_{x}\tau^{2}-J_{y})(2f_{k}^{2\dagger}f_{k}^{2}-1)+J_{z}(\tau^{1}+\tau^{2})\\
&+(K+\tilde{K}\tau^{1}\tau^{2})\bigg(f_{k}^{1}f_{-k}^{2}-f_{k}^{1\dagger}f_{k}^{2}+f_{k}^{1}f_{k}^{2\dagger}-f_{k}^{1\dagger}f_{-k}^{2\dagger}\bigg)\\
&+(K+\tilde{K}\tau^{1}\tau^{2})\bigg(f_{k}^{2}f_{-k}^{1}e^{ik}-f_{k}^{2\dagger}f_{k}^{1}e^{-ik}+f_{k}^{2}f_{k}^{1\dagger}e^{ik}-f_{k}^{2\dagger}f_{-k}^{1\dagger}e^{-ik}\bigg).
\end{aligned}
\end{eqnarray}
To simplify the notations we introduce coupling constants $J_{1}=J_{x}\tau^{1}-J_{y}$, $J_{2}=J_{x}\tau^{2}-J_{y}$, $K_{1}=K+\tilde{K}\tau^{1}\tau^{2}$. The Hamiltonian can then be written in a matrix form, 
\begin{equation}
\label{fIsingmatrix}
\mathcal{H}_{v}^{(d)}=\sum_{k}\left(\begin{array}{cccc}
f_{k}^{1\dagger}&f_{k}^{2\dagger}&f_{-k}^{1}&f_{-k}^{2}
\end{array}\right)\hat{\mathbf{H}}_{k}\left(\begin{array}{c}
f_{k}^{1}\\f_{k}^{2}\\f_{-k}^{1\dagger}\\f_{-k}^{2\dagger}
\end{array}\right)+J_{z}(\tau^{1}+\tau^{2}),
\end{equation}
the matrix $\hat{\mathbf{H}}_{k}$ is given by
\begin{equation}
\hat{\mathbf{H}}_{k}=\left(\begin{array}{cccc}
J_{1}&-\frac{1}{2}K_{1}(1+e^{ik})&0&-\frac{1}{2}K_{1}(1-e^{ik})\\
-\frac{1}{2}K_{1}(1+e^{-ik})&J_{2}&\frac{1}{2}K_{1}(1-e^{-ik})&0\\
0&\frac{1}{2}K_{1}(1-e^{ik})&-J_{1}&\frac{1}{2}K_{1}(1+e^{ik})\\
-\frac{1}{2}K_{1}(1-e^{-ik})&0&\frac{1}{2}K_{1}(1+e^{-ik})&-J_{2}
\end{array}\right)
\end{equation}
The phase diagram is thus obtained by comparing ground-state energies of the fermionic Hamiltonian evaluated on a grid of $k$-space with different values of $\tau^{1}$ and $\tau^{2}$.

\begin{figure}
\includegraphics[width=0.45\textwidth]{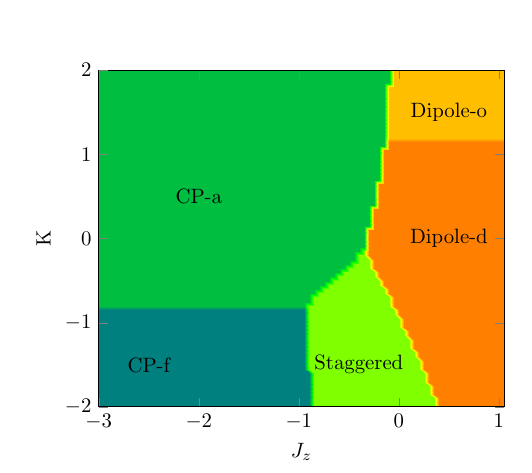}
\includegraphics[width=0.45\textwidth]{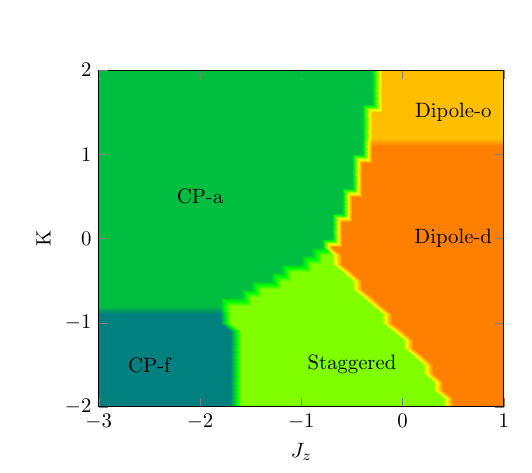}
\caption{Comparison between phase diagrams of the vertical ladder obtained in two different ways with $\tilde{K}=0.8$, $J_{x}=J_{y}=1$. Left: from exact calculation of fermionic ground states; Right: from numerical diagonalization on a system of 16 spins (open boundary condition).}
\label{figcompare}
\end{figure}

The vertical ladder can be mapped into a free fermion model and thus enjoys exact solvability. For the horizontal ladder, the only way to obtain the phase diagram is by diagonalizing the spin Hamiltonian on a finite lattice. The vertical ladder is thus an ideal platform to study the finite-size effect. To this end, we adopt another way to obtain the phase diagram of the vertical ladder, namely by diagonalization of the spin Hamiltonian (Eq. (4) in the main text) on a system of 16 spins. The phase diagrams of the vertical ladder obtained by the two methods are compared in Fig. \ref{figcompare}, from which the finite-size effect of the direct diagonalization of spin Hamiltonian is evident. In particular the phase boundaries of the ``staggered" and ``CP" phases are moved to the left.

Following the main text, we write down the model with plaquette terms which results in XX and YY coupling of the first-order spin $\boldsymbol{S}$,
\begin{equation}
    \mathcal{H}_{v}^{(d)'}=\sum_{n=1}^{2N-1}K_{p}^{x}\sigma_{2n-1}^{y}\sigma_{2n}^{y}\sigma_{2n+1}^{y}\sigma_{2n+2}^{y}+K_{p}^{y}\sigma_{2n-1}^{y}\sigma_{2n}^{x}\sigma_{2n+1}^{x}\sigma_{2n+2}^{y}\rightarrow \sum_{n=1}^{2N-1}K_{p}^{x}S_{n}^{x}S_{n+1}^{x}+K_{p}^{y}S_{n}^{y}S_{n+1}^{y}.
\end{equation}
Added to the Hamiltonian \eqref{SmodeloneS} we obtain the full Hamiltonian 
\begin{eqnarray}
\label{Smodelonefull}
    \begin{aligned}
        \mathcal{H}_{v}^{(d)}=&\sum_{n=1}^{2N}\left(J_{x}\sigma_{2n-1}^{x}\sigma_{2n}^{x}+J_{y}\sigma_{2n-1}^{y}\sigma_{2n}^{y}+J_{z}\sigma_{2n-1}^{z}\sigma_{2n}^{z}\right)+\\&\sum_{n=1}^{2N-1}\left(K\sigma_{2n}^{z}\sigma_{2n+1}^{z}+\tilde{K}\sigma_{2n-1}^{z}\sigma_{2n+2}^{z}+K_{p}^{x}\sigma_{2n-1}^{y}\sigma_{2n}^{y}\sigma_{2n+1}^{y}\sigma_{2n+2}^{y}+K_{p}^{y}\sigma_{2n-1}^{y}\sigma_{2n}^{x}\sigma_{2n+1}^{x}\sigma_{2n+2}^{y}\right),
    \end{aligned}
\end{eqnarray}
which would translate into the general XYZ model in transverse field for the first-order spin,
\begin{eqnarray}
\label{SmodeloneXYZ}
    \begin{aligned}
        \mathcal{H}_{v}^{(d)}=&\sum_{n=1}^{2N}\left[(J_{y}-J_{x}\tau_{n})S_{n}^{x}+J_{z}\tau_{n}\right]\\&+\sum_{n=1}^{2N-1}(K+\tilde{K}\tau_{n}\tau_{n+1})S_{n}^{z}S_{n+1}^{z}+K_{p}^{x}S_{n}^{x}S_{n+1}^{x}+K_{p}^{y}S_{n}^{y}S_{n+1}^{y}.
    \end{aligned}
\end{eqnarray}
Such a model supports richer phase diagrams than the transverse-field Ising model.

\subsection{The horizontal ladder}

For the horizontal ladder, the lattice geometry is taken as half-infinite and has an extra spin site for definition purpose, we name this site 0 as shown in Fig. \ref{figSladder}. The spin Hamiltonian can be written explicitly as
\begin{equation}
\mathcal{H}_{h}^{(d)}=\sum_{n=0}\left[J_{x}\sigma_{2n}^{x}\sigma_{2n+1}^{x}+J_{y}\sigma_{2n}^{y}\sigma_{2n+1}^{y}+J_{z}\sigma_{2n}^{z}\sigma_{2n+1}^{z}\right]+\sum_{n=1}\left[K\sigma_{2n-1}^{z}\sigma_{2n}^{z}+\tilde{K}\sigma_{2n-1}^{z}\sigma_{2n+2}^{z}\right].
\end{equation}
In the following we take the coupling on bond $(0,1)$ to be zero so that the spin on site 0 does not couple to the ladder. Using JW transformation \eqref{JWsupplemental} with the starting point of string operator $s=0$, we have 
\begin{eqnarray}
\begin{aligned}
\mathcal{H}_{h}^{(d)}=&\sum_{n=0}\left[J_{x}(-)\eta_{2n}^{\beta}\eta_{2n+1}^{\alpha}\eta_{2n}^{\alpha}\eta_{2n+1}^{\beta}+J_{y}(-i)\eta_{2n+1}^{\alpha}\eta_{2n}^{\beta}+J_{z}i\eta_{2n+1}^{\beta}\eta_{2n}^{\alpha}\right]\\&+\sum_{n=1}\left[Ki\eta_{2n}^{\beta}\eta_{2n-1}^{\alpha}+\tilde{K}(-i)\eta_{2n+2}^{\beta}\eta_{2n-1}^{\alpha}\eta_{2n}^{\alpha}\eta_{2n}^{\beta}\eta_{2n+1}^{\alpha}\eta_{2n+1}^{\beta}\right].
\end{aligned}
\end{eqnarray}
The model has conserved quantities $\tau_{n}=\sigma_{2n}^{z}\sigma_{2n+1}^{z}=-i\eta_{2n}^{\alpha}\eta_{2n+1}^{\beta}$, with which the Hamiltonian is written as
\begin{equation}
\label{Hhorizontalf}
\mathcal{H}_{h}^{(d)}=\sum_{n=0}\left[(J_{y}-J_{x}\tau_{n})i\eta_{2n}^{\beta}\eta_{2n+1}^{\alpha}+J_{z}\tau_{n}\right]+\sum_{n=1}\left[K(-i)\eta_{2n-1}^{\alpha}\eta_{2n}^{\beta}+\tilde{K}(-\tau_{n})\eta_{2n-1}^{\alpha}\eta_{2n}^{\beta}\eta_{2n+1}^{\alpha}\eta_{2n+2}^{\beta}\right].
\end{equation}
Then we group together sites $(2n,2n+1)$ with $n=0,1,2,\cdots$, and redefine $\eta_{2n}^{\beta}\rightarrow \tilde{\eta}_{n}^{\beta}$ and $\eta_{2n+1}^{\alpha}\rightarrow \tilde{\eta}_{n}^{\alpha}$. The first-order spin $\boldsymbol{S}_{n}$ is introduced for each pair by a new set of JW transformation 
\begin{equation}
\tilde{\eta}_{n}^{\alpha}=S_{n}^{y}\prod_{j=0}^{n-1}(-S_{j}^{x}),\qquad \tilde{\eta}_{n}^{\beta}=S_{n}^{z}\prod_{j=0}^{n-1}(-S_{j}^{x}),\qquad i\tilde{\eta}_{n}^{\beta}\tilde{\eta}_{n}^{\alpha}=S_{n}^{x},
\end{equation}
using which the Hamiltonian (\ref{Hhorizontalf}) becomes 
\begin{equation}
\mathcal{H}_{h}^{(d)}=\sum_{n=0}\left[(J_{y}-J_{x}\tau_{n})S_{n}^{x}+J_{z}\tau_{n}\right]+\sum_{n=1}\left[KS_{n-1}^{z}S_{n}^{z}+\tilde{K}\tau_{n}S_{n-1}^{z}S_{n+1}^{z}\right].
\end{equation}
With static distribution of $\{\tau_{n}\}$ the Hamiltonian involves Ising interaction of second-nearest neighbours, making it the ANNNI model. 

We then look for relations between the first-order spin $\boldsymbol{S}$ and the zeroth-order spin $\boldsymbol{\sigma}$. Noting that 
$S_{n}^{z}=\eta_{2n}^{\beta}\prod_{j=0}^{n-1}(-i\eta_{2j}^{\beta}\eta_{2j+1}^{\alpha})$ and $\sigma_{2n}^{z}=\eta_{2n}^{\beta}\prod_{j=0}^{2n-1}(i\eta_{j}^{\alpha}\eta_{j}^{\beta})$, we have 
\begin{eqnarray}
\begin{aligned}
\sigma_{2n}^{z}=&\eta_{2n}^{\beta}(-i\eta_{0}^{\beta}\eta_{0}^{\alpha})(-i\eta_{1}^{\beta}\eta_{1}^{\alpha})(-i\eta_{2}^{\beta}\eta_{2}^{\alpha})(-i\eta_{3}^{\beta}\eta_{3}^{\alpha})\cdots (-i\eta_{2n-1}^{\beta}\eta_{2n-1}^{\alpha})\\
=&S_{n}^{z}\tau_{0}\tau_{1}\cdots\tau_{n-1}.
\end{aligned}
\end{eqnarray}
Using $\tau_{n}=\sigma_{2n}^{z}\sigma_{2n+1}^{z}=-i\eta_{2n}^{\alpha}\eta_{2n+1}^{\beta}$, this becomes the first part of the relation $S_{n}^{z}=\prod_{j=0}^{2n}\sigma_{j}^{z}$. The second part of the relation $S_{n}^{x}=\sigma_{2n}^{y}\sigma_{2n+1}^{y}$ holds for the same reason as the vertical ladder, see Eq. \eqref{Srelationsigma}. 

Here the dipole parity can also be defined, namely $\mathcal{I}=\prod_{n=0}S_{n}^{z}=\prod_{n=0}\prod_{j=0}^{2n}\sigma_{j}^{z}$, satisfying $\{S_{n}^{x},\mathcal{I}\}=0$. Explicitly the dipole parity can be written as $\mathcal{I}=\prod_{j}(\sigma_{j}^{z})^{N-x_{j}}$, where $x_{j}$ is the effective coordinate of the spin site $j$, which coincides with the $x$-coordinates of the sites in the horizontal geometry shown in Fig. 1 of the main text.

\section*{The quadrupole model and beyond}

Following the main text, here we present detailed definition of the phases of the quadrupole model. The configurations of one plaquette with four spins $\sigma_{4n-3}$ to $\sigma_{4n}$ in the quadrupole model can be classified by two first-order $\mathbb{Z}_{2}$ quantities $\tau_{2n-1}=\sigma_{4n-3}^{z}\sigma_{4n-2}^{z}$ and $\tau_{2n}=\sigma_{4n-1}^{z}\sigma_{4n}^{z}$ and one second-order $\mathbb{Z}_{2}$ quantity $T_{n}=S_{2n-1}^{z}S_{2n}^{z}=\sigma_{4n-2}^{z}\sigma_{4n-1}^{z}$. We name the configurations with two parts. The first part reflects the values of the two $\tau$; if they both equal to $-1$, we call it ``dipole"; if they both equal to $+1$, we call it ``charge-pair (CP)"; if they are not equal, we call it ``staggered". The second part reflects the value of $T$, for $T=-1$ it is ``dipole" and for $T=+1$ it is ``charge-pair (CP)". One special example is for $\tau_{2n-1}=\tau_{2n}=T_{n}=-1$, then the configuration is ``dipole-dipole" which is naturally ``quadrupole". In the table below we list all the possible configurations determined by $\tau$ and $T$, and give pictorial examples for each case. 

\begin{tabular}{|c|c|c|c|c|c|}
	\hline
	configuration &$\tau_{2n-1}$&$\tau_{2n}$&$T_{n}$& example& abbreviation\\
	\hline
	quadrupole&-1&-1&-1&\begin{tikzpicture}
	\draw[very thick,green](0.1,0) -- (0.4,0);
	\draw[very thick,green](0.5,0.1) -- (0.5,0.4);
	\draw[very thick,green](0,0.1) -- (0,0.4);
	\draw[very thick,green](0.1,0.5) -- (0.4,0.5);
	\filldraw (0,0) circle [radius=0.1];
	\draw (0,0.5) circle [radius=0.1];
	\draw (0.5,0) circle [radius=0.1];
	\filldraw (0.5,0.5) circle [radius=0.1];
	\end{tikzpicture}&Q\\
	\hline
	dipole-CP&-1&-1&+1&\begin{tikzpicture}
	\draw[very thick,green](0.1,0) -- (0.4,0);
	\draw[very thick,green](0.5,0.1) -- (0.5,0.4);
	\draw[very thick,green](0,0.1) -- (0,0.4);
	\draw[very thick,green](0.1,0.5) -- (0.4,0.5);
	\filldraw (0,0) circle [radius=0.1];
	\draw (0,0.5) circle [radius=0.1];
	\draw (0.5,0.5) circle [radius=0.1];
	\filldraw (0.5,0) circle [radius=0.1];
	\end{tikzpicture}&D-CP\\
	\hline
	staggered-dipole&-1&+1&-1&\begin{tikzpicture}
	\draw[very thick,green](0.1,0) -- (0.4,0);
	\draw[very thick,green](0.5,0.1) -- (0.5,0.4);
	\draw[very thick,green](0,0.1) -- (0,0.4);
	\draw[very thick,green](0.1,0.5) -- (0.4,0.5);
	\filldraw (0,0) circle [radius=0.1];
	\draw (0,0.5) circle [radius=0.1];
	\filldraw (0.5,0.5) circle [radius=0.1];
	\filldraw (0.5,0) circle [radius=0.1];
	\end{tikzpicture}&S-D\\
	\hline
	staggered-CP&-1&+1&+1&\begin{tikzpicture}
	\draw[very thick,green](0.1,0) -- (0.4,0);
	\draw[very thick,green](0.5,0.1) -- (0.5,0.4);
	\draw[very thick,green](0,0.1) -- (0,0.4);
	\draw[very thick,green](0.1,0.5) -- (0.4,0.5);
	\filldraw (0,0) circle [radius=0.1];
	\draw (0,0.5) circle [radius=0.1];
	\draw (0.5,0.5) circle [radius=0.1];
	\draw (0.5,0) circle [radius=0.1];
	\end{tikzpicture}&S-CP\\
	\hline
	staggered-dipole&+1&-1&-1&\begin{tikzpicture}
	\draw[very thick,green](0.1,0) -- (0.4,0);
	\draw[very thick,green](0.5,0.1) -- (0.5,0.4);
	\draw[very thick,green](0,0.1) -- (0,0.4);
	\draw[very thick,green](0.1,0.5) -- (0.4,0.5);
	\filldraw (0,0) circle [radius=0.1];
	\filldraw (0,0.5) circle [radius=0.1];
	\draw (0.5,0.5) circle [radius=0.1];
	\filldraw (0.5,0) circle [radius=0.1];
	\end{tikzpicture}&S-D\\
	\hline
	staggered-CP&+1&-1&+1&\begin{tikzpicture}
	\draw[very thick,green](0.1,0) -- (0.4,0);
	\draw[very thick,green](0.5,0.1) -- (0.5,0.4);
	\draw[very thick,green](0,0.1) -- (0,0.4);
	\draw[very thick,green](0.1,0.5) -- (0.4,0.5);
	\filldraw (0,0) circle [radius=0.1];
	\filldraw (0,0.5) circle [radius=0.1];
	\filldraw (0.5,0.5) circle [radius=0.1];
	\draw (0.5,0) circle [radius=0.1];
	\end{tikzpicture}&S-CP\\
	\hline
	CP-dipole&+1&+1&-1&\begin{tikzpicture}
	\draw[very thick,green](0.1,0) -- (0.4,0);
	\draw[very thick,green](0.5,0.1) -- (0.5,0.4);
	\draw[very thick,green](0,0.1) -- (0,0.4);
	\draw[very thick,green](0.1,0.5) -- (0.4,0.5);
	\filldraw (0,0) circle [radius=0.1];
	\filldraw (0,0.5) circle [radius=0.1];
	\draw (0.5,0.5) circle [radius=0.1];
	\draw (0.5,0) circle [radius=0.1];
	\end{tikzpicture}&CP-D\\
	\hline
	CP-CP&+1&+1&+1&\begin{tikzpicture}
	\draw[very thick,green](0.1,0) -- (0.4,0);
	\draw[very thick,green](0.5,0.1) -- (0.5,0.4);
	\draw[very thick,green](0,0.1) -- (0,0.4);
	\draw[very thick,green](0.1,0.5) -- (0.4,0.5);
	\draw (0,0) circle [radius=0.1];
	\draw (0,0.5) circle [radius=0.1];
	\draw (0.5,0.5) circle [radius=0.1];
	\draw (0.5,0) circle [radius=0.1];
	\end{tikzpicture}&CP-CP\\
	\hline
\end{tabular}

In order to study the phase diagram of the quadrupole model, we note that the original model $\mathcal{H}^{(q)}$ has degeneracy between ``quadrupole" phase and ``CP-dipole" phase as well as ``dipole-CP" phase and ``CP-CP" phase. To lift such degeneracy we add to the Hamiltonian an extra term of Ising ZZ coupling on horizontal bonds, and set $\tilde{J}^{z}=J^{z}$. The original Hamiltonian of the zeroth-order spin now becomes
\begin{eqnarray}
\label{higherorderHmodified}
    \begin{aligned}
        \mathcal{H}^{(q)}=&\sum_{n=1}^{N}\left(J_{p}^{x}\prod_{j=4n-3}^{4n}\sigma_{j}^{x}+J_{p}^{y}\prod_{j=4n-3}^{4n}\sigma_{j}^{y}\right)+\sum_{n=1}^{N}J^{z}(\sigma_{4n-2}^{z}\sigma_{4n-1}^{z}+\sigma_{4n-3}^{z}\sigma_{4n}^{z})+\mathcal{J}^{z}(\sigma_{4n-3}^{z}\sigma_{4n-2}^{z}+\sigma_{4n-1}^{z}\sigma_{4n}^{z})\\&+\sum_{n=1}^{N-1}(K\sigma_{4n}^{z}\sigma_{4n+1}^{z}+\tilde{K}\sigma_{4n-1}^{z}\sigma_{4n+2}^{z})
    \end{aligned}
\end{eqnarray}
In terms of the second-order spin, the modified Hamiltonian \eqref{higherorderHmodified} can be written as
\begin{equation}
\label{higherorderIsingmodified}
    \mathcal{H}^{(q)}=\sum_{n=1}^{N}-(J_{p}^{x}\tau_{2n-1}\tau_{2n}+J_{p}^{y})T_{n}\mathbb{S}_{n}^{x}+\sum_{n=1}^{N}\left[J^{z}(1+\tau_{2n-1}\tau_{2n})T_{n}+\mathcal{J}^{z}(\tau_{2n-1}+\tau_{2n})\right]+\sum_{n=1}^{N-1}(K+\tilde{K}\tau_{2n}\tau_{2n+1})\mathbb{S}_{n}^{z}\mathbb{S}_{n+1}^{z}.
\end{equation}
To obtain the phase diagram, we take the parameters $J_{p}^{x}=J_{p}^{y}=1$, $\mathcal{J}^{z}=0.1$ and $\tilde{K}=0.8$. This Ising Hamiltonian can be turned into a fermionic model like \eqref{fIsing} in the vertical ladder. We assume that the ground-state of the model respects translational symmetry such that the unit cell contains two green plaquettes (see Fig. \ref{figquadrupole}). We are left with six conserved quantities for every unit cell which we call $\tau_{1}$ to $\tau_{4}$ and $T_{1}$, $T_{2}$, as shown below. 

\begin{figure}
\includegraphics[width=0.45\textwidth]{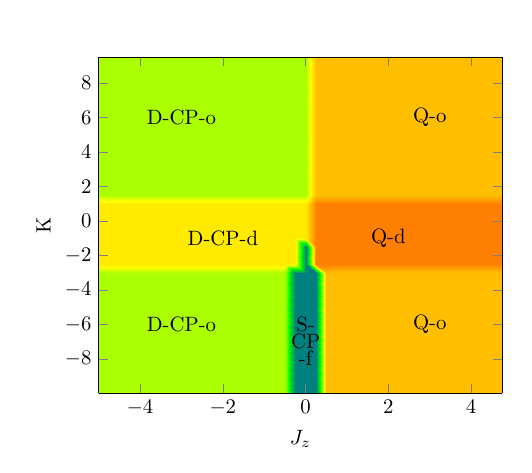}
\includegraphics[width=0.45\textwidth]{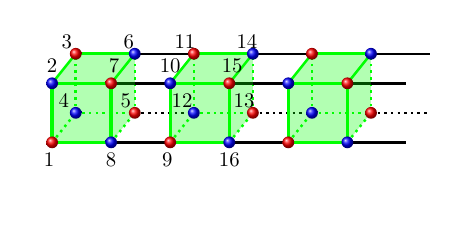}
\caption{Left: phase diagram of the quadrupole model \eqref{higherorderHmodified} with $J_{p}^{x}=J_{p}^{y}=1$, $\mathcal{J}=0.1$, $\tilde{K}=0.8$. Right: lattice structure of the octopole model with numbering of the sites.}
\label{figSthree}
\end{figure}

\begin{tikzpicture}
\foreach \i in {0,2}
{
	\filldraw[fill=green!20,draw=green]
	  (\i,0)--(\i,1)--(\i+1,1)--(\i+1,0);
	\draw[very thick](\i+1,0)--(\i+2,0);
	\draw[very thick,dashed](\i+1,1)--(\i+2,1);
	\draw[ultra thick,green](\i,0)--(\i,1);
	\draw[ultra thick,green](\i+1,0)--(\i+1,1);
	\draw[ultra thick,green](\i,1)--(\i+1,1);
	\draw[ultra thick,dashed,green](\i,0)--(\i+1,0);
	\filldraw[fill=red] (\i,0) circle [radius=0.15];
	\filldraw[fill=blue] (\i,1) circle [radius=0.15]; 
	\filldraw[fill=red] (\i+1,1) circle [radius=0.15];
	\filldraw[fill=blue] (\i+1,0) circle [radius=0.15]; 
}

\node at (0.5,1.3) {$T_{1}$};
\node at (-0.2,0.5) {$\tau_{1}$};
\node at (2.5,1.3) {$T_{2}$};
\node at (1.2,0.5) {$\tau_{2}$};
\node at (1.8,0.5) {$\tau_{3}$};
\node at (3.2,0.5) {$\tau_{4}$};
\end{tikzpicture}
\newline
Following \eqref{fIsingmatrix} we define quantities 
\begin{equation}
    J_{1}=(J_{p}^{x}\tau_{1}\tau_{2}+J_{p}^{y})T_{1}, \qquad J_{2}=(J_{p}^{x}\tau_{3}\tau_{4}+J_{p}^{y})T_{2},\qquad K_{1}=K+\tilde{K}\tau_{2}\tau_{3},\qquad K_{2}=K+\tilde{K}\tau_{1}\tau_{4}.
\end{equation}
The rest follows the vertical ladder case. The phases are further classified by the phases of the transverse-field Ising model. Specifically, for the ``quadrupole" phase and the ``dipole-CP" phase, if $-2.8<K<1.2$, the system is in the disordered phase of $\mathbb{S}$, and the phases are labelled by ``Q-d" and ``D-CP-d" respectively. Otherwise the system is in ordered phase, labelled by ``Q-o" and ``D-CP-o". For ``staggered-dipole" phase, if $K>0.8$, the system is in antiferromagnetic phase for $\mathbb{S}$, labelled by ``S-D-a"; for $K<0.8$ the system is in ferromagnetic phase for $\mathbb{S}$, labelled by ``S-D-f". The phase diagram is given in Fig. \ref{figSthree}.

Generalization to octopole models is possible by adding another layer to the ladder geometry. A zig-zag string is still possible to be defined linking all the sites together. We plot the lattice structure in Fig. \ref{figSthree}.

\section*{Simulating the dipole model using electric field controlled Rydberg atoms}

Here we discuss in detail the possible simulation of the simplest dipole model on vertical ladder using electric-field controlled Rydberg atoms. For such simulation, the Rydberg states of atoms, for example $^{87}\text{Rb}$, are prepared; the atoms are aligned in ladder geometry using optical tweezers. A weak electric field is applied to control the states and our choice of the pseudospin up and down are the Stark-shifted states $n_{\uparrow}S_{\frac{1}{2}}$ and $n_{\downarrow}S_{\frac{1}{2}}$. In other words, we identify
\begin{equation}
    |\uparrow\rangle\rightarrow |\widetilde{n_{\uparrow}S}\rangle,\qquad |\downarrow\rangle\rightarrow |\widetilde{n_{\downarrow}S}\rangle,
\end{equation}
in which the tilde sign is used to denote Stark-shifted states. The distance between nearest neighbor interacting atoms is denoted by $R$. Without electric field, the atoms interact with van der Waals (vdW) interaction, which is second-order and in general proportional to $1/R^{6}$. With electric field, the Stark shifted states have non-vanishing matrix elements over the dipole operator, resulting in a direct dipole-dipole interaction, which is proportional to $1/R^{3}$ and depends on the direction of the electric field.  

\begin{figure}
\includegraphics[width=0.55\textwidth]{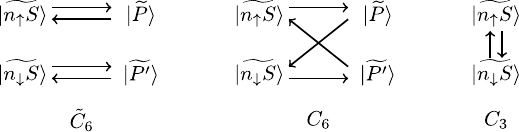}
\caption{Origin of the interactions between the atomic states. vdW interactions $C_{6}$ and $\tilde{C}_{6}$ originate from second-order perturbation mediated by $|P\rangle$ states. Dipole-dipole interaction of the Stark shifted states $C_{3}$ is first-order and directly couples the two states.}
\label{figSryd}
\end{figure}

Here we neglect the interaction beyond nearest neighboring atoms and focus on two neighboring atoms. The Hilbert space contains four states $(|\uparrow\uparrow\rangle,|\uparrow\downarrow\rangle,|\downarrow\uparrow\rangle,|\downarrow\downarrow\rangle)^{T}$. Within this subspace the effective Hamiltonian for these two neighboring spins reads
\begin{equation}
\label{Hefftwosites}
    \mathcal{H}_{\text{eff}}=\left(\begin{array}{cccc}
        \frac{C_{6}^{\uparrow}}{R^{6}} & 0&0&0 \\
         0&\frac{\tilde{C}_{6}}{R^{6}}&V(R,\theta)&0\\
         0&V(R,\theta)&\frac{\tilde{C}_{6}}{R^{6}}&0\\
         0&0&0&\frac{C_{6}^{\downarrow}}{R^{6}}
    \end{array}\right),
\end{equation}
where $V(R,\theta)$ takes into account indirect second-order vdW interaction and the first-order dipole-dipole interaction
\begin{equation}
\label{VRtheta}
    V(R,\theta)=\frac{C_{6}}{R^{6}}+\frac{C_{3}}{R^{3}}(1-3\cos^{2}\theta).
\end{equation}
The origins of these couplings between atomic states are illustrated in Fig. \ref{figSryd}. The effective interaction between the two neighboring sites leads to a many-body spin Hamiltonian for the entire system, whose Hamiltonian is 
\begin{equation}
    \mathcal{H}=\sum_{\langle ij\rangle}\left[J_{z}^{ij}\sigma_{i}^{z}\sigma_{j}^{z}+J_{xy}^{ij}(\sigma_{i}^{x}\sigma_{j}^{x}+\sigma_{i}^{y}\sigma_{j}^{y})\right]+h\sum_{i}\sigma_{i}^{z}.
\end{equation}
The coupling strength can be obtained from the two-sites effective Hamiltonian \eqref{Hefftwosites}, namely
\begin{equation}
    J_{z}=\frac{C_{6}^{\uparrow}+C_{6}^{\downarrow}-2\tilde{C}_{6}}{4R^{6}},\qquad J_{xy}=\frac{V(R,\theta)}{2},\qquad h=\frac{(C_{6}^{\uparrow}-C_{6}^{\downarrow})}{4R^{6}}.
\end{equation}
In the ladder geometry, the angle between the electric field and the horizontal are different from that of the vertical bonds and so $J_{xy}$ takes different values on vertical bonds and horizontal bonds. Our goal is to tune the distance $R$ and angle $\theta$ such that 
\begin{equation}
    V(R,\theta)=0,\qquad \text{for the horizontal bonds}.
\end{equation}
By using microwave field, we also tune the energies of the involved atomic states such that $C_{6}^{\uparrow}=C_{6}^{\downarrow}$ and thus $h\rightarrow 0$. In this way we have obtained the dipole model on the vertical ladder.

\section*{Dynamical structural factor}

The zero-temperature dynamical structural factor can be used to distinguish the phases of the ladder models. By definition, it is given by 
\begin{equation}
\label{DSFchain}
D^{\alpha\beta}_{0}(q,\omega)=\frac{1}{N}\sum_{j,l=1}^{N}\int_{-\infty}^{\infty}dt e^{iql+i\omega t}\langle S_{j}^{\alpha}(t)S_{j-l}^{\beta}(0)\rangle
\end{equation}
for spin chain. For the 2D spin ladder we are considering, the zero-temperature dynamical structural factor is given by
\begin{equation}
\label{DSFladder}
D^{zz}(q_{x},q_{y},\omega)=\frac{1}{2N}\sum_{j,l=1}^{N}\sum_{y,y'=0,1}\int dt e^{i\omega t}e^{iq_{x}l+iq_{y}(y-y')}\langle S_{j+l,y}^{z}(t)S_{j,y'}^{z}(0)\rangle.
\end{equation}
Here we denote the structural factor by $D$ instead of the standard $S$ to avoid confusion with the spin.

Firstly we consider the vertical ladder, for which the sum over $y,y'$ in \eqref{DSFladder} can be done before the integration and gives
\begin{equation}
D^{zz}(q_{x},q_{y},\omega)=\frac{1}{2N}\sum_{j,l=1}^{N}\int dt e^{i\omega t+iq_{x}l}\left[\sum_{y,y'=0,1}e^{iq_{y}(y-y')}\langle S_{j+l,y}^{z}(t)S_{j,y'}^{z}(0)\rangle\right].
\end{equation}
If the system is in the {\it dipole phase}, namely $S_{i,1}^{z}\equiv -S_{i,0}^{z}$, the dynamical structural factor is 
\begin{equation}
D^{zz}(q_{x},q_{y},\omega)=(1-\cos q_{y})\frac{1}{N}\sum_{j,l=1}^{N}\int dt e^{i\omega t+iq_{x}l}\langle S_{j+l,0}^{z}(t)S_{j,0}^{z}(0)\rangle=(1-\cos q_{y})D_{0}^{zz}(q_{x},\omega).
\end{equation}
If it is in the {\it charge-pair phase}, namely $S_{i,1}^{z}\equiv S_{i,0}^{z}$, the dynamical structural factor is 
\begin{equation}
D^{zz}(q_{x},q_{y},\omega)=(1+\cos q_{y})\frac{1}{N}\sum_{j,l=1}^{N}\int dt e^{i\omega t+iq_{x}l}\langle S_{j+l,0}^{z}(t)S_{j,0}^{z}(0)\rangle=(1+\cos q_{y})D_{0}^{zz}(q_{x},\omega).
\end{equation}
To summarize, if $S_{i,1}^{z}\equiv \pm S_{i,0}^{z}$, the dynamical structural factor is $D^{zz}(q_{x},q_{y},\omega)=(1\pm\cos q_{y})D_{0}^{zz}(q_{x},\omega)$, for which $D_{0}^{zz}(q_{x},\omega)$ is the dynamical structural factor for the Ising chain defined in \eqref{DSFchain}.

The zero-temperature dynamical structural factor for the horizontal ladder can be obtained similarly. By definition we label the unit cells of the lattices by its coordinates $(x,y)$; inside each unit cell, the two sites are labeled by 0 and 1. The dynamical structural factor \eqref{DSFladder} can be written as
\begin{eqnarray}
\begin{aligned}
D^{zz}(q_{x},q_{y},\omega)=&\frac{1}{2N}\sum_{x,x'}\sum_{y,y'}\int dt e^{i\omega t}e^{iq_{y}(y-y')}e^{iq_{x}(x-x')}\big[\langle S_{x,0,y}^{z}(t)S^{z}_{x',0,y'}(0)\rangle\\&+\langle S_{x,0,y}^{z}(t)S_{x',1,y'}^{z}(0)\rangle e^{-iq_{x}}+\langle S_{x,1,y}^{z}(t)S_{x',0,y'}^{z}(0)\rangle e^{iq_{x}}+\langle S_{x,1,y}^{z}(t)S_{x',1,y'}^{z}(0)\rangle\big].
\end{aligned}
\end{eqnarray}
For the {\it dipole phase} of the horizontal ladder, $S_{x,1,y}\equiv -S_{x,0,y}$ for all unit cells. So the dynamical structural factor reads
\begin{eqnarray}
\begin{aligned}
D^{zz}(q_{x},q_{y},\omega)&=(1-\cos q_{x})\bigg[\frac{1}{N}\sum_{x,x'}\sum_{y,y'}\int dt e^{i\omega t}e^{iq_{y}(y-y')+iq_{x}(x-x')}\langle S^{z}_{x,0,y}(t)S^{z}_{x',0,y'}(0)\rangle\bigg]\\
&=(1-\cos q_{x})\tilde{D}^{zz}(q_{x},q_{y},\omega).
\end{aligned}
\end{eqnarray}
In the {\it charge-pair phase} of the horizontal ladder, $S_{x,1,y}\equiv S_{x,0,y}$ for all unit cells. So the dynamical structural factor is given by
\begin{eqnarray}
\begin{aligned}
D^{zz}(q_{x},q_{y},\omega)&=(1+\cos q_{x})\bigg[\frac{1}{N}\sum_{x,x'}\sum_{y,y'}\int dt e^{i\omega t}e^{iq_{x}(x-x')+iq_{y}(y-y')}\langle S^{z}_{x,0,y}(t)S^{z}_{x',0,y'}(0)\rangle\bigg]\\
&=(1+\cos q_{x})\tilde{D}^{zz}(q_{x},q_{y},\omega).
\end{aligned}
\end{eqnarray}
In the above, $\tilde{D}^{zz}(q_{x},q_{y},\omega)$ represents the zero-temperature dynamical structural factor for a zig-zag ANNNI model. To summarize, if $S_{x,1,y}^{z}\equiv \pm S_{x,0,y}^{z}$, the zero-temperature dynamical structural factor for the horizontal ladder is $D^{zz}(q_{x},q_{y},\omega)=(1\pm\cos q_{x})\tilde{D}^{zz}(q_{x},q_{y},\omega)$. As discussed in the main text, the dipole phase and the charge-pair phase are distinguished experimentally by the {\it zeros} of the dynamical structural factor.

\end{document}